\documentclass[epsf,11pt]{article}

\usepackage{epsfig}

\newcommand\nc{\newcommand}

\nc{\textsize}[2]{
\textwidth=#1
\oddsidemargin=8.5in
\addtolength\oddsidemargin{-2in}
\addtolength\oddsidemargin{-\textwidth}
\divide\oddsidemargin by2
\evensidemargin=\oddsidemargin
\textheight=#2
\topmargin=11in
\addtolength\topmargin{-3in}
\addtolength\topmargin{-\textheight}
\divide\topmargin by2
}

\newtheorem{theorem}{Theorem}[section]
\newtheorem{lemma}[theorem]{Lemma}
\newtheorem{cor}[theorem]{Corollary}

\newtheorem{ques}[theorem]{Question}

\nc{\crl}[2]{\begin{cor}\label{crl:#1} #2 \end{cor}}
\nc{\dfn}[2]{\begin{defn}\label{def:#1} #2 \end{defn}}
\nc{\lem}[2]{\begin{lemma}\label{lem:#1} #2 \end{lemma}}
\nc{\thm}[2]{\begin{theorem}\label{thm:#1} #2 \end{theorem}}
\nc{\que}[2]{\begin{ques}\label{que:#1} #2 \end{ques}}

\nc{\fig}[4]{\begin{figure}[hbt]
\centerline{
\epsfysize=#2in
\epsffile{#4}
}
\caption{#3}
\label{fig:#1}
\end{figure}}

\nc{\refc}[1]{Corollary~\ref{crl:#1}}
\nc{\refd}[1]{Definition~\ref{def:#1}}
\nc{\reff}[1]{Figure~\ref{fig:#1}}
\nc{\refl}[1]{Lemma~\ref{lem:#1}}
\nc{\reft}[1]{Theorem~\ref{thm:#1}}
\nc{\refb}[1]{Problem~\ref{pro:#1}}

\def\qed{\rule{1.5mm}{3mm}}

\nc{\proof}[1]{ {\bf Proof.} #1 \hfill \qed\par}

\nc{\pf}[1]{{\bf Proof.} #1 \hfill \qed\par}
\textsize{5.2in}{8.5in}

\title{\Large \bf Alternating Path and Coloured Clustering}
\author{{\large CAI Leizhen\thanks{Email: lcai@cse.cuhk.edu.hk, Partially supported by CUHK
Direct Grant 4055069} \hspace{5pt} and LEUNG On Yin\thanks{Email: clp01234544@gmail.com}}  \\ \\
\normalsize \sl Department of Computer Science and Engineering \\
\normalsize \sl  The Chinese University of Hong Kong \\
\normalsize \sl Shatin, New Territories, Hong Kong SAR, China} 
\date{\today}

\begin{document}
\maketitle

\begin{abstract}
In the {\sc Coloured Clustering} problem, we wish to colour vertices
of an edge coloured graph to produce as many stable edges as possible, i.e., 
edges with the same colour as their ends.
In this paper, we reveal that the problem is in fact a maximum subgraph
problem concerning monochromatic subgraphs and alternating paths, and 
demonstrate the usefulness of such connection in studying these problems.

We obtain a faster algorithm to solve the problem for edge-bicoloured graphs
by reducing the problem to a minimum cut problem.
On the other hand, we push the NP-completeness of the problem to edge-tricoloured 
planar bipartite graphs of maximum degree four.
Furthermore, we also give FPT algorithms for the problem when we take the numbers 
of stable edges and unstable edges, respectively, as parameters.\\
\end{abstract}

\section{Introduction}

The following {\sc Coloured Clustering} problem has been proposed recently by 
Angel et al.~\cite{Angel} in connection with the classical correlation 
clustering problem~\cite{Bansal}: Compute a vertex colouring of 
an edge-coloured graph $G$ to produce as many {\em stable edges}
as possible, i.e., edges with the same colour as their ends.
As observed by Ageev and Kononov~\cite{Ageev}, the problem contains
the classical maximum matching problem as a special case as the two problems
coincide when all edges have different colours in $G$.

In this paper we will reveal that {\sc Coloured Clustering}, despite its 
definition by vertex partition, is in fact the following maximum subgraph problem 
in disguise: 
find a largest subgraph where every vertex has one colour for its incident edges 
({\sc Vertex-Monochromatic Subgraph}), 
or, equivalently, delete fewest edges to destroy all alternating paths 
({\sc Alternating Path Removal}).
This multiple points of view gives us a better understanding of these problems,
and is quite useful in studying them.

We are mainly interested in algorithmic issues of {\sc Coloured Clustering}, 
and will consider polynomial-time algorithms, NP-completeness, and also FPT algorithms
for the two natural parameters from the subgraph point of view: 
numbers of edges inside ({\em stable edges}) and outside ({\em unstable edges}), 
respectively, the solution subgraph.

\subsection{Main results}

We now summarize our main results for {\sc Coloured Clustering}, 
where $m$ and $n$, respectively, are numbers of edges and vertices in $G$.
These results can be translated directly into corresponding results for 
{\sc Vertex-Monochromatic Subgraph} and {\sc Alternating Path Removal}.

\begin{itemize}
\item We obtain an $O(m^{3/2}\log n)$-time algorithm for edge-bicoloured graphs $G$ by 
a reduction to the classical minimum cut problem,  
which improves the $O(m^{3/2}n)$-time algorithm of Angel et al.~\cite{Angel} 
based on independent sets in bipartite graphs.
We also give linear-time algorithms for the special case when $G$ is a complete graph 
(see \S4).

\item We push the NP-completeness of the problem to edge-tricoloured planar 
bipartite graphs of maximum degree four (see \S5).

\item We derive FPT algorithms for the problem when we take the numbers 
of stable edges and unstable edges, respectively, as parameter $k$,
which is uncommon for most problems parameterized in this way.
Furthermore, we obtain a kernel with at most $4k$ vertices and $2k^2+k$ edges
for the latter problem (see \S6).
\end{itemize}

\subsection{Related work}

Both monochromatic subgraphs and alternating paths are at least half-century old, 
and there is a huge number of papers in the literature dealing with 
them graph theoretically~\cite{Bang,Kano}.
However, we are not aware of any work on these two subjects 
that is directly related to the algorithmic problems we study in this paper.

In the literature, papers by Angel et al.~\cite{Angel} and Ageev and Kononov~\cite{Ageev}
seem to be the only work that directly study {\sc Coloured Clustering}.
Angel et al. obtain an LP-based $1/e^2$-approximation algorithm for the problem in general, 
which is improved to a 7/23-approximation algorithm by Ageev and Kononov.
Angel et al. also give a polynomial-time algorithm for the problem on edge-bicoloured
graphs by a reduction to the maximum independent set problem on bipartite graphs,
but show the NP-completeness of the problem for edge-tricoloured bipartite graphs.

\section{Definitions}

An {\em edge-coloured graph} $G$ is a simple graph where each edge $e$ has a unique
colour $\psi(e) \in  \{1, \dots, t\}$ for some positive integer $t$.
We say that $G$ is {\em edge-bicoloured} if $t = 2$, and {\em edge-tricoloured}
if $t =3$.
Unless specified otherwise, we use $m$ and $n$, respectively,
for the numbers of edges and vertices of $G$.

A vertex $v$ is {\em colourful} if its incident edges have at least 
two different colours, and {\em monochromatic} otherwise.
A subgraph $H$ of $G$ is {\em vertex-monochromatic} if all vertices in $H$
are monochromatic vertices of $H$, and {\em edge-monochromatic} if all edges in $H$ 
have the same colour. 

A {\em conflict pair} is a pair of adjacent edges of different colours,
and an {\em alternating path} is a simple path where every pair of consecutive
edges forms a conflict pair.
The {\em edge-conflict graph} of $G$, denoted $X(G)$, is an uncoloured graph
where each vertex represents an edge of $G$ 
and each edge corresponds to a conflict pair in $G$.

A {\em vertex colouring} $f$ of $G$ assigns to each vertex $v$ of $G$ a colour 
$f(v) \in  \{1, \dots, t\}$ for some positive integer $t$. 
For a vertex colouring $f$ of $G$, an edge $uv$ is {\em stable} if its colour
$\psi(uv) = f(u) = f(v)$, and {\em unstable} otherwise.

Angel et al.~\cite{Angel} have recently proposed the following problem,
which is in fact, as we will see shortly, the problem of finding the largest
vertex-monochromatic subgraph in $G$ in disguise (see \refl{equal}).

\begin{quote}
{\sc Coloured Clustering} \\
{\sc Input}: Edge coloured graph $G$ and positive integer $k$. \\
{\sc Question}: Is there a vertex colouring of $G$ that produces at least $k$ stable edges?
\end{quote}

The following problem is concerned with purging conflict-pairs
(equivalently, alternating paths) by edge deletion, and is
the complementary problem of {\sc Coloured Clustering} (see \refc{same}).

\begin{quote}
{\sc Conflict-Pair Removal} \\
{\sc Input}: Edge coloured graph $G$ and positive integer $k$. \\
{\sc Question}: Does $G$ contain at most $k$ edges $E'$ such that $G - E'$ 
contains no conflict pair?
\end{quote}

\section{Basic properties}

Although {\sc Coloured Clustering} is defined by vertex partition 
(i.e., vertex colouring), it is in fact a maximum subgraph problem in disguise.
To see this, we first observe the following equivalent properties for
edge-coloured graphs.  

\lem{equal}{The following statements are equivalent for any edge-coloured graph $G$:\\
{\rm (a).} $G$ is vertex-monochromatic.\\
{\rm (b).} Every component of $G$ is edge-monochromatic.\\
{\rm (c).} $G$ has no alternating path. \\
{\rm (d).} $G$ has no conflict pair.
}
\pf{
The equivalence between (a) and (b) is obvious, and so is the equivalence 
between (c) and (d) as a conflict pair is itself an alternating path. 
Furthermore, it is again obvious that $G$ contains a conflict pair 
if and only if $G$ has a colourful vertex, and therefore (a) and (d) are equivalent.
It follows that the four statements are indeed equivalent.
}

Observe that for any vertex colouring of $G$, the subgraph formed by stable edges
is vertex-monochromatic, and hence {\sc Coloured Clustering}
is actually equivalent to finding a largest vertex-monochromatic subgraph,
which in turn is equivalent to deleting fewest edges to destroy all conflict pairs.
This gives us the following complementary relation between {\sc Coloured Clustering}
and {\sc Conflict-Pair Removal}.

\crl{same}{
There is vertex colouring for $G$ that produces at least $k$ stable edges if and only if 
$G$ contains at most $m-k$ edges $E'$ such that $G-E'$ has no conflict pair.
}

The bilateral relation between {\sc Coloured Clustering} and {\sc Conflict-Pair Removal} 
is akin to that between the classical {\sc Independent Set} and {\sc Vertex Cover}.
In fact, the former two problems become exactly the latter two in the edge-conflict 
graph $X(G)$ of $G$ (see \reft{conflict-graph}).

It is very useful to view {\sc Coloured Clustering} as an edge deletion problem, 
instead of a vertex partition problem, which often makes things easier. 
For instance, it becomes straightforward to obtain the following result 
of Angel et al.~\cite{Angel} for $X(G)$. 

\thm{conflict-graph}{{\rm [Angel et al.\cite{Angel}]}
An edge-coloured graph $G$ admits a vertex colouring that produces at least $k$ 
stable edges if and only if the edge-conflict graph $X(G)$ of $G$ has 
an independent set of size at least $k$.
}
\pf{
By \refc{same}, the former statement is equivalent to deleting at most 
$m-k$ edge to obtain a graph without conflict pair, which is the same as $X(G)$
has a vertex cover of size at most $m-k$, and hence $X(G)$ has an independent 
set of size at least $k$.
}

\section{Algorithms for edge-bicoloured graphs}

Although {\sc Coloured Clustering} is NP-complete for edge-tricoloured graphs~\cite{Angel},
Angel et al.~\cite{Angel} have obtained an $O(m^{3/2}n)$-time algorithm for the problem
on edge-bicoloured graphs $G$ by reducing it to the maximum independent set problem on
bipartite graphs $X(G)$. 
In this section, we will give a faster $O(m^{3/2}\log n)$-time algorithm
by considering {\sc Conflict-Pair Removal}, which leads us to a simple reduction
to a minimum cut problem.
We also give a linear-time algorithm for the problem on edge-bicoloured complete graphs.

\subsection{Faster algorithm}

One bottleneck of the algorithm of Angel et al. lies in the size of the edge-conflict
graph $X(G)$ which contains $O(m)$ vertices and $O(mn)$ edges.
Here we use a different approach of reduction to construct a digraph $G'$ with only 
$O(m)$ vertices and edges, and then solve an equivalent minimum cut problem on $G'$
to solve our problem.

Let $G = (V, E)$ be an edge-bicoloured graph with colours $\{1,2\}$,
and consider {\sc Conflict-Pair Removal}.
Our idea is to transform every conflict pair in $G$ into an $(s, t)$-path 
in a digraph $G'$ with source $s$ and sink $t$.
For this purpose, we construct digraph $G'$ from $G$ as follows
(see \reff{mincut} for an example of the construction):
\begin{enumerate}
\item Take graph $G$ and add two new vertices --- source $s$ and sink $t$.
\item For each edge $v_iv_j$ of $G$, create a new vertex $v_{ij}$ to represent edge $v_iv_j$.
If $v_iv_j$ has colour 1 then replace it by two edges $v_{ij}v_i$ and $v_{ij}v_j$ 
and add edge $sv_{ij}$.
Otherwise replace the edge by two edges $v_iv_{ij}$ and $v_jv_{ij}$ 
and add edge $v_{ij}t$.
\end{enumerate}

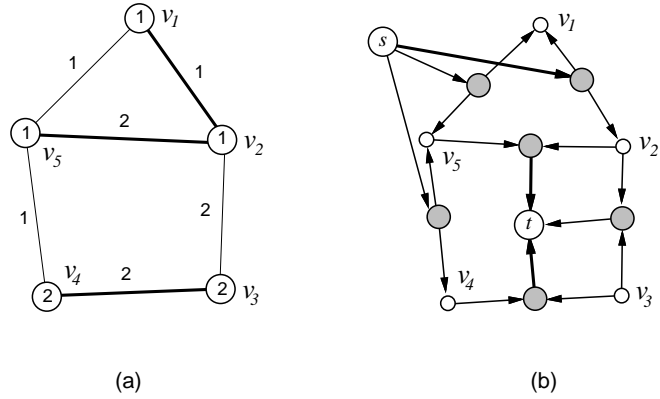
\fig{mincut}{2.1}{Digraph $G'$ from graph $G$, where shaded vertices in $G'$ 
correspond to edges in $G$ and thick edges indicate corresponding solution edges.}{mincut.fig}

An {\em $(s, t)$-cut} in $G'$ is a set of edges whose deletion disconnects 
sink $t$ from source $s$.
Let $v_iv_j$ and $v_iv_{j'}$ be an arbitrary conflict pair of $G$.
Without loss of generality, we may assume that $v_iv_j$ has colour 1
and $v_iv_{j'}$ has colour 2.
By the construction of $G'$, there is a unique $(s, t)$-path
\[ P(j,i,j')  = sv_{ij}, v_{ij}v_i, v_iv_{ij'}, v_{ij'}t \]
in $G'$ that goes through vertices $v_{ij}$ and $v_{ij'}$.
For convenience, we refer to edges $sv_{ij}$ and $v_{ij'}t$
as {\em external edges} and the other two edges as {\em middle edges}.
Edges $v_iv_j$ and $v_iv_{j'}$ of $G$ correspond to external edges
$sv_{ij}$ and $v_{ij'}t$, respectively, in $G'$.
We also call an $(s, t)$-cut a {\em normal cut} if
the cut contains no middle edge of any $P(j,i,j')$.

\lem{mincut}{
Let $E'$ be a set of edges in $G$.
Then $G - E'$ contains no conflict pair if and only if corresponding edges
of $E'$ in $G'$ form a normal $(s, t)$-cut of $G'$.
}
\pf{
There is a one-to-one correspondence between conflict pair $v_iv_j$ 
and $v_iv_{j'}$ in $G$ and external edges $sv_{ij}$ and $v_{ij'}t$ in $G'$ 
in such a way that the conflict pair is destroyed if and only if 
the $(s, t)$-path $P(j,i,j')$ is disconnected. This clearly implies the lemma.
}

The above lemma enables us to solve {\sc Conflict-Pair Removal} on edge-bicoloured
graphs by reducing it to the minimum cut problem, which yields a faster algorithm.

\thm{bicoloured}{{\sc Conflict-Pair Removal} for edge-bicoloured graphs $G$ can be
solved in $O(m^{3/2}\log n)$.
}
\pf{
By \refl{mincut}, we can reduce our problem on $G$ to the minimum normal $(s, t)$-cut
problem on digraph $G'$.
Observe that for any $(s, t)$-cut, we can always replace a middle edge by 
an external edge without increasing the size of the cut.
Therefore we need only solve the minimum $(s, t)$-cut problem on digraph $G'$,
which can be accomplished by the maximum flow algorithm of Goldberg and Rao~\cite{Gold}.
 
For the running time of the algorithm, we first note that $G'$ contains $N = m + n + 2$ vertices
and $M = 3m$ edges, and can be constructed in $O(m + n)$ time.
Since every edge of $G'$ has capacity 1, Goldberg and Rao's algorithm takes  
$O(\min(N^{2/3}, M^{1/2})M \log N^2/M)$ time,
which gives us $O(m^{3/2}\log n)$ time as $M, N = O(m)$.
}

\crl{bicoloured}{{\sc Coloured Clustering} for edge-bicoloured graphs $G$ can be
solved in $O(m^{3/2}\log n)$ time.
}

\subsection{Complete graphs} 

We now turn to the special case of {\sc Coloured Clustering} when $G = (V, E)$ 
is an edge-bicoloured complete graph, and present a linear-time algorithm.
Let $f$ be a vertex-2-colouring of $G$ that colours vertices $V_1$ by colour 1
and vertices $V_2$ by colour 2.
For a vertex $v$, let $d_1(v)$ be the number of edges of colour 1 incident with $v$.
Let $m_1$ be the number of edges with colour 1.
We can completely determine the number of stable edges produced by $f$ as follows.

\lem{clique}{
For a vertex-$2$-colouring $f$ of an edge-bicoloured complete graph $G$, the number $S_f$
of stable edges produced by $f$ equals
\[\sum_{v \in V_1} d_1(v) +  {|V_2| \choose 2} - m_1.\]
}
\pf{Let $A$ and $B$ be numbers of edges of colour 1 
in $G[V_1]$ and $G[V_2]$ respectively.
By the definition of stable edges, we have
\[ S_f = A + ({|V_2| \choose 2} - B) \]
as $G[V_2]$ is a complete graph.
On the other hand,  $B  = m_1 - C$,
where $C$ is the number of edges of colour 1 covered by vertices $V_1$.
Therefore
\[ S_f = A + {|V_2| \choose 2} + C - m_1, \]
and the lemma follows from the fact that  
$A + C = \sum_{v \in V_1} d_1(v)$.
}

With the formula in the above lemma, we can easily and efficiently 
solve {\sc Coloured Clustering} for edge-bicoloured complete graphs.

\crl{clique}{
{\sc Coloured Clustering} can be solve in $O(n^2)$ for edge-bicoloured complete graphs.
}
\pf{
From \refl{clique}, we see that once we fix the size of $V_1$ to be $k$,
$S_f$ is maximized when we choose $k$ vertices $v$ with largest $d_1(v)$
as vertices in $V_1$.
Therefore we can compute the maximum value of $S_f$ for each $0 \le k \le n$,
and find an optimal vertex-2-colouring for $G$.
The whole process clearly takes $O(n^2)$ time as we can first sort vertices 
according to $d_1(v)$.
}

We can also use a similar idea to solve {\sc Coloured Clustering} in $O(n^2)$ time
for edge-bicoloured complete bipartite graphs, which will appear in our full paper.

\section{NP-completeness}

Angel et al.~\cite{Angel} have shown the NP-completeness of 
{\sc Coloured Clustering} for edge-tricoloured bipartite graphs.
In this section, we further push the intractability of the problem
to edge-tricoloured planar bipartite graphs of bounded degree.
Recall that a vertex colouring is {\em proper} if the two ends 
of every edge receive different colours.

\thm{planar}{
{\sc Coloured Clustering} is NP-complete for edge-tricoloured 
planar bipartite graphs of maximum degree four.
}
\pf{
Garey, Johnson and Stockmeyer~\cite{Garey} proved the NP-completeness of {\sc Independent Set} 
on cubic planar graphs, and we give a reduction from this restricted case of 
{\sc Independent Set} to our problem.
For an arbitrary cubic planar graph $G = (V, E)$ with $V = \{ v_1, \dots, v_n \}$, 
we construct an edge-tricoloured planar bipartite graph $G' = (V', E')$ of 
maximum degree four as follows:

\begin{enumerate}
\item Compute a proper vertex 3-colouring $\psi$ of $G$.

\item For each edge $v_iv_j \in E$, subdivide it by a new vertex $v_{ij}$
(i.e., replace edge $v_iv_j$ by two edges $v_iv_{ij}$ and $v_{ij}v_j$),
and colour edges $v_iv_{ij}$ and $v_{ij}v_j$ by $\psi(v_i)$ and $\psi(v_j)$ respectively.

\item For each vertex $v_i \in V$, add a new vertex $v^*_i$ and edge $v_iv_i^*$,
and colour edge $v_iv_i^*$ by a colour in $\{1,2,3\}$ different from $\psi(v_i)$.
\end{enumerate}

It is clear that $G'$ is an edge-tricoloured planar bipartite graphs of maximum degree four.
By Brooks' Theorem, every cubic graph except $K_4$ admits a proper vertex 3-colouring,
and we can use an algorithm of Lov\'{a}sz~\cite{Lovasz} to compute a proper 
vertex 3-colouring of a cubic graph in linear time. 
Therefore, the above construction of $G'$ takes polynomial time.
We claim that $G$ has an independent set of size $k$ if and only if $G'$
admits a vertex colouring that produces $k + |E|$ stable edges.

Suppose that $G$ contains an independent set $I$ of size $k$.
We define a vertex-3-colouring $f$ of $G'$ as follows:
\begin{enumerate}
\item For each vertex $v_i^* \in V^*$, set $f(v_i^*)$ to be the colour of edge $v_iv_i^*$. 

\item For each vertex $v_i \in V$, set $f(v_i)$ to be the colour of edge $v_iv^*_i$ if $v_i \in I$
and $f(v_i) = \psi(v_i)$ otherwise.

\item For each vertex $v_{ij}$, set $f(v_{ij})$ to be $\psi(v_i)$ if $v_i \not\in I$ 
and $\psi(v_j)$ otherwise.
\end{enumerate}

Clearly, $f$ produces $k$ stable edges $v_iv_i^*$ after Step 2. 
For any vertex $v_{ij}$, since $I$ contains at most one of $v_i$ and $v_j$,
exactly one of $v_iv_{ij}$ and $v_{ij}v_j$ becomes a stable edge after Step 3.
Therefore $f$ produces $k+ |E|$ stable edges for $G'$.

Conversely, call each edge $v_iv_i^*$ an {\em outside edge},
and let $f'$ be a vertex 3-colouring of $G'$ that produces $k + |E|$ stable edges
and also minimizes the number of outside edges among these stable edges.
Let 
\[ I = \{ v_i: \mbox{edge $v_iv^*_i$ is stable} \}.\]
For every vertex $v_{ij}$ in $G'$, since edges $v_iv_{ij}$ and $v_{ij}v_j$ have different colours,
at most one of these two edges is a stable edge for any vertex colouring of $G'$.
It follows that at least $k$ stable edges are formed by outside edges and hence
$I$ contains at least $k$ vertices.

We claim that $I$ is an independent set of $G$.
Suppose to the contrary that for some vertices $v_i, v_j \in I$, $v_iv_j$ is an edge of $G$.
First we note that amongst all edges incident with $v_i$,  
$v_iv^*_i$ is the only stable edge under $f'$ 
as $f'(v_i) \not= \psi(v_i)$, and similar situation holds for all edges incident with $v_j$.
In particular, neither $v_iv_{ij}$ nor $v_{ij}v_j$ is a stable edge.
We now recolour both vertices $v_i$ and $v_{ij}$ by the colour of edge $v_iv_{ij}$
(note that $v_{ij}$ may have received that colour already under $f'$) 
to obtain a new vertex 3-colouring $f''$ (see \reff{npc} for an example of the situation).

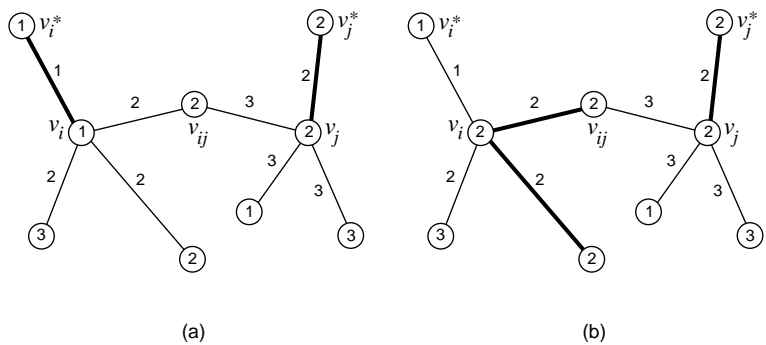
\fig{npc}{1.8}{(a) Situation under vertex colouring $f'$.  
(b) Situation after recolouring vertices $v_i$ and $v_{ij}$. 
Stable edges are indicated by thick edges.}{npc.fig}

Comparing with colouring $f'$, this new colouring $f''$ reduces one stable edge 
(namely, edge $v_iv_i^*$), but produces a new stable edge $v_iv^*_i$ 
(and probably also other new stable edges).
Therefore $f''$ produces at least $k + |E|$ stable edges that contains 
one less outside edges than $f'$, contradicting the choice of $f'$.
This contradiction implies that $I$ is indeed an independent set of $G$ with 
at least $k$ vertices, and hence the theorem holds.
}

\crl{planar}{
{\sc Conflict-Pair Removal} is NP-complete for edge-tricoloured
planar bipartite graphs of maximum degree four.
}

\section{FPT algorithms}

We now turn to the parameterized complexity of {\sc Coloured Clustering},
and give FPT algorithms for the problem with respect to both the number 
of stable edges and the number of unstable edges as parameter $k$.
This is quite interesting as it is uncommon for a problem to admit
FPT algorithms both ways when parameterized in this manner.

\subsection{Stable edges}

First we take the number of stable edges produced by a vertex colouring 
of $G$ as parameter $k$, and use {\sc Coloured Cluster}$[k]$ to denote this
parameterized problem.  
We will give an FPT algorithm that uses random partition in the spirit of 
the colour coding method of Alon, Yuster and Zwick~\cite{Alon}, which 
implies an FPT algorithm for {\sc Independent Set}$[k]$ in edge-conflict graphs.
Note that if the number $t$ of colours in $G$ is a constant, 
then the problem is trivially solved in FPT time as it contains a trivial kernel 
with at most $kt$ edges and hence $2kt$ vertices.
Also note that the problem is not as easy as it looks, for it contains
the maximum matching problem as a special case when all edges have different colours.

Our idea is to randomly partition vertices of $G$ into $k$ parts $V_1, \dots, V_k$
in a hope that a $k$-solution consists of $k_i$ stable edges, where $\sum_{i=1}^k k_i = k$,
in each $G[V_i]$. Indeed we have a good chance to succeed in this way.

\begin{quote}
{\bf Algorithm} Coloured-Clustering$[k]$\\
Randomly partition vertices $V$ of $G$ into $V_1,\cdots,V_k$.\\
Compute the most frequently used colour $c_i$ for  each $G[V_i]$. \\
Colour all vertices in $V_i$ by $c_i$.\\
\end{quote}

\lem{chance}{For any yes-instance of \/ {\sc Coloured Cluster}$[k]$,
the vertex colouring constructed by {\bf Algorithm} {\rm Coloured-Clustering}$[k]$ 
has probability at least $k^{-2k}$ to produce at least $k$ stable edges.
}
\pf{Consider a vertex colouring of $G$ that produces at least $k$ stable edges $E'$,
which clearly have at most $k$ different colours.
Let $E'_i$ be edges in $E'$ of colour $c_i$ and let $k_i = |E'_i|$ for $1 \le i \le k$.
We estimate the probability that all edges of $E'_i$ lie in $G[V_i]$.
A vertex has probability $k^{-1}$ to be in $V_i$, and hence 
the above event happens with probability at least $k^{-2k_i}$ 
as $E'_i$ contains at most $2k_i$ vertices. 
It follows that, with probability at least 
\[ k^{- \sum_{i=1}^k 2k_i} = k^{-2k}, \]
all edges of each $E'_i$ lie entirely inside $G[V_i]$.
Therefore each $G[V_i]$ contains at least $k_i$ edges of same colour $c_i$,
which can be made stable by colouring all vertices in $G[V_i]$ by colour $c_i$.
It follows that the algorithm produces at least $k$ stable edges with probability $k^{-2k}$.
}

The algorithm runs in $O(k^{2k}(m+n))$ expected time, and can be made into a 
deterministic FPT algorithm by standard derandomization with a family of
perfect hashing functions.

\thm{stable}{
{\sc Coloured Cluster}$[k]$ can be solved in {\rm FPT} time.
}

\subsection{Unstable edges}

Now we take the number of unstable edges in a vertex colouring as parameter $k$,
and use {\sc Conflict-Pair Removal}$[k]$ to denote this parameterized problem.
By a result of Angel et al.~\cite{Angel}, the problem is equivalent to finding a
vertex cover of size $k$ in the edge-conflict graph $X(G)$ of $G$, 
and hence admits an FPT algorithm by transforming it to the $k$-vertex cover problem 
in $X(G)$.
However the time for the transformation takes $O(mn)$ time as $X(G)$ contains $O(m)$
vertices and $O(mn)$ edges, and the total time for the algorithm takes 
$O(mn + 1.2783^k)$ time. 
Here we combine kernelization with weighted vertex cover to obtain
an improved algorithm with running time $O(m+n+1.2783^k)$.

To start with, we construct in linear time the following edge-coloured weighted graph $G^*$, 
called {\em condensed graph}, by representing monochromatic vertices of one colour by a single
vertex, and then parallel edges between two vertices by a single weighted edge.
See \reff{condense} for an example of the construction.

{\bf Step 1.} For each colour $c$, contract all monochromatic vertices of colour $c$ 
into a single vertex $v_c$.
 
{\bf Step 2.} For each pair of adjacent vertices, if there is only one edge between them,
then set the weight of the edge to 1, otherwise replace all parallel edges between them 
by a single edge of the same colour\footnote{All such parallel edges have the same colour 
as they correspond to edges between a vertex and monochromatic vertices of the same colour.}
and set its weight to be the number of replaced parallel edges.

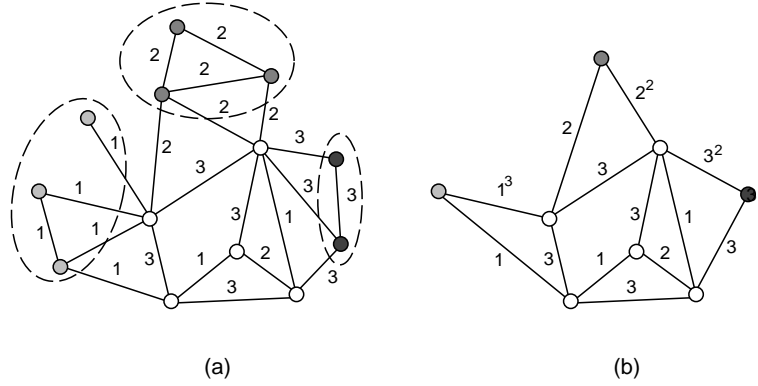
\fig{condense}{2}{(a) Edge-coloured graph $G$ where each dashed ellipse 
indicates monochromatic vertices of same colour.
(b) The condensed graph $G^*$ of $G$ where an edge of weight more than 1 
has its weight as the superscript of its colour.}{condense.fig}

It turns out that the clustering problem on $G$ is equivalent to a weighted version of the problem 
on the condensed graph $G^*$.

\lem{condense}{
Graph $G$ has at most $k$ unstable edges if and only if 
$G^*$ has unstable edges of total weight at most $k$.
}
\pf{By the construction of $G^*$, we have the following correspondence 
between edges in $G$ and $G^*$: 
every edge in $G$ between two colourful vertices remains so in $G^*$, and 
for any colourful vertex $v$, all edges between $v$ and monochromatic vertices 
of colour $c$ correspond to edge $vv_c$ in $G^*$.
Also all monochromatic vertices in $G^*$ form an independent set.

Now suppose that $G$ has a vertex colouring $f$ that produces $k$ unstable edges.
Without loss of generality, we may assume that every monochromatic vertex $v$ in $G$
has its own colour as $f(v)$ since this will not increase unstable edges.
For this $f$, we have a natural vertex colouring $f^*$ for $G^*$: the colour of each vertex
retains its colour under $f$.
It is obvious that an edge in $G^*$ between two colourful vertices is an unstable edge under
$f^*$ 
if and only if it is an unstable edge in $G$ under $f$.
For edges $E_c(v)$ in $G$ between a colourful vertex $v$ and monochromatic vertices 
with colour $c$, either all edges in $E_c(v)$ are stable or all are unstable
as $f$ colours all these monochromatic vertices by colour $c$, 
implying that all edges in $E_c(v)$ are unstable under $f$ if and only if 
$vv_c$ is unstable under $f^*$.
Therefore $f^*$ produces unstable edges of total weight $k$ in $G^*$.

Conversely, suppose that $G^*$ contains a set $U$ of unstable edges of total weight $k$, and
let $U'$ be the corresponding $k$ edges in $G$.
Clearly $G - U'$ contains no conflict pair, and hence $G$ has at most $k$ unstable edges.
}

Further to the above lemma, $G^*$ can be regarded as a kernel as its size
is bounded by a function of $k$ whenever $G$ has at most $k$ unstable edges.
Note that the bounds in the following lemma are tight.

\lem{kernel}{
If $G$ has at most $k$ unstable edges, then $G^*$ has at most $4k$ vertices and $2k^2+k$ edges.
}
\pf{Let $[C, M]$ be the cut that partitions the vertices of $G$ 
into colourful vertices $C$ and monochromatic vertices $M$.
Let $A$ be the set of unstable edges inside $G[C]$, and $B$ the set of unstable edges across the cut.
Observe that each edge of $A$ is incident with at most two vertices of $C$, 
and each edge of $B$ is incident with one vertex of $C$.
Furthermore, every colourful vertex is incident with at least one unstable edge.
Therefore $|C| \le 2|A|+|B| \le 2k$.

Now consider the condensed graph $G^*$, and note that the cut $[C, M]$ corresponds to the cut $[C,
M^*]$
for monochromatic vertices $M^*$ of $G^*$.
Furthermore, $A$ consists of unstable edges inside $G^*[C]$, and
$B$ corresponds to unstable edges $B^*$ across $[C, M^*]$ and $|B^*| \le |B|$.

In $G^*$, every vertex in $C$ is incident with at most one stable edge in  $[C, M^*]$.
Therefore $[C, M^*]$ contains at most
\[|C| + |B^*| \le (2|A|+|B|)+|B^*| \le 2(|A|+|B|) = 2k\]
edges, and hence $M^*$ contains at most $2k$ vertices.
It follows that $G^*$ contains at most $4k$ vertices, and at most 
${2k \choose 2}+2k = 2k^2+k$ edges as $G^*[M^*]$ is edgeless.
}

With \refl{condense} and \refl{kernel} in hand, we obtain the following FPT algorithm
for {\sc Conflict-Pair Removal}$[k]$

\begin{tabbing}
{\bf Algorithm} Conflict-Pair-Removal$[k]$ \\

Construct the condensed graph $G^*$ from $G$; \\
{\bf if} $G^*$ \= contains more than $4k$ vertices or $2k^2+k$ edges \\
	\> {\bf then return} ``No'' and {\bf halt}; \\
Construct the edge-conflict $X(G^*)$ of $G^*$;\\
{\bf if} $X(G^*)$ has a vertex cover of weight at most $k$ \\
	\> {\bf then return} ``Yes''\\
	\> {\bf else return} ``No''.\\
\end{tabbing}

\thm{}{{\sc  Conflict-Pair Removal}$[k]$ can be solved in $O(m+n+1.2783^k)$ time.
}
\pf{The correctness of the algorithm follows from \refl{condense} and \refl{kernel}, 
and we analyze the running time of the algorithm.
The construction of the condensed graph $G^*$ clearly takes $O(m+n)$ time,
and the construction of the edge-conflict graph $X(G^*)$ takes $O(k^3)$ time 
as $G^*$ contains $O(k)$ vertices and $O(k^2)$ edges.
Note that $X(G^*)$ contains $O(k^2)$ vertices and $O(k^3)$ edges.
Since it takes $O(kn + 1.2783^k)$ to solve the weighted vertex cover problem~\cite{Chen, Nied}, 
it takes $O(k^3 + 1.2783^k) = O(1.2783^k)$ to solve the problem for $G^*$,
and hence the overall time is $O(m + n + 1.2783^k)$.
}

\section{Concluding remarks}

We have revealed that {\sc Coloured Clustering}, a vertex partition problem, 
is in fact subgraph problems {\sc Vertex-Monochromatic Subgraph} and
{\sc Alternating Path Removal} in disguise, and demonstrated the usefulness of 
this multiple points of view in studying these problems.
Indeed, our improved algorithm for edge-bicoloured graphs and FPT algorithms 
for general edge-coloured graphs have benefited a lot from 
the perspective of {\sc Conflict-Pair Removal}.
We now briefly discuss a few open problems in the language of monochromatic subgraphs
and alternating paths for readers to ponder. \\ 

\noindent {\bf Question 1.} {\em For edge-bicoloured graphs, is there a faster algorithm 
for deleting fewest edges to obtain a vertex-monochromatic subgraph?
}

There seems to be a good chance to solve the problem faster than our algorithm,
and one possible approach is to reduce the number of vertices in the reduction 
to minimum cut from the current $O(m)$ to $O(n)$.\\

\noindent {\bf Question 2.} {\em For {\sc Conflict-Pair Removal} on general 
edge-coloured graphs, is there an $r$-approximation algorithm for some constant $r < 2$?}

The problem admits a simple 2-approximation algorithm through its connection with
{\sc Vertex Cover}, and seems easier than the latter problem.
It is possible that we can do better for the problem, perhaps through ILP relaxation. \\

\noindent {\bf Question 3.} {\em For edge-coloured graphs, 
does the problem of finding a vertex-monochromatic subgraph with at least $k$ edges 
admit a polynomial kernel?
}

The above problem is appealing for its connection with the classical maximum matching 
problem.
On one hand, we may use a maximum matching of $G$ as a starting point for 
a polynomial kernel; and on the other hand if we can obtain a polynomial kernel of $G$
in $o(m\sqrt{n})$ time, we may use the kernel to speed up maximum matching algorithms.
Of course, it may be the case that the problem admits no polynomial kernel unless
NP $\subseteq$ coNP/poly.\\

\noindent {\bf Question 4.} {\em For edge-coloured graphs, 
is there an FPT algorithm for the problem of destroying 
all alternating cycles by deleting at most $k$ edges?
}

Although the definition of the problem resembles that of {\sc Alternating Path Removal},
the problem seems much more difficult as the problem does not have a finite forbidden
structure like conflict pair for the latter problem.
We note that the problem is NP-complete by a simple reduction from {\sc Vertex Cover}.

%\section*{Acknowledgement}

\bibliography{ref}
{}

\bibliographystyle{plain}

\end{document}